# SAT problem and Limit of Solomonoff's inductive reasoning theory

PAN Feng

(College of Information Science and Technology, Donghua Univesity, Shanghai 201620)

**ABSTRACT:** This paper explores the Boolean Satisfiability Problem (SAT) in the context of Kolmogorov complexity theory. We present three versions of the distinguishability problem—Boolean formulas, Turing machines, and quantum systems—each focused on distinguishing between two Bernoulli distributions induced by these computational models. A reduction is provided that establishes the equivalence between the Boolean formula version of the program output statistical prediction problem and the #SAT problem. Furthermore, we apply Solomonoff's inductive reasoning theory, revealing its limitations: the only "algorithm" capable of determining the output of any shortest program is the program itself, and any other algorithms are computationally indistinguishable from a universal computer, based on the coding theorem. The quantum version of this problem introduces a unique algorithm based on statistical distance and distinguishability, reflecting a fundamental limit in quantum mechanics. Finally, the potential equivalence of Kolmogorov complexity between circuit models and Turing machines may have significant implications for the NP vs P problem. We also investigate the nature of short programs corresponding to exponentially long bit sequences that can be compressed, revealing that these programs inherently contain loops that grow exponentially.

## 1. Introduction

Around 1971, Cook [1] and Levin [2] independently discovered the concept of NP-completeness (NPC) and provided examples of combinatorial problems that are NPC. Soon after, Karp [3] demonstrated that NPC problems were widespread and that many practically important problems fell into this category. Today, thousands of problems from nearly all branches of science have been proven to be NPC problems [4,5]. The existence of an efficient polynomial-time algorithm for any NPC problem is equivalent to the question of whether NP equals P. In this context, we sought to analyze the complexity of the Boolean Satisfiability problem (SAT) through the lens of Kolmogorov complexity theory, leveraging the connection between these two areas.

Any Boolean function $f: \{0,1\}^n \to \{0,1\}$ can be represented using basic gates such as AND, OR, NOT, and FANOUT, which together constitute a universal gate set for classical computing. It follows that there are $2^{2^n}$ possible functions within this function space. By using gates to express calculations, it has been shown that certain functions require exponential size of gates to perform calculations.

**Theorem 1.1** (Existence of hard functions). For every $n>1$, there exists a function $f: \{0,1\}^n \to \{0,1\}$, that cannot be computed by a circuit $C$ of size $2^n/(10n)$.

Further analysis leaded to a stronger conclusion than theorem 1.1[4]: not only are there hard functions which cannot be calculated by any circuit of size at most $2^n/(10n)$, but in fact the vast majority of all the possible functions are hard.

A Boolean formula $\varphi$ involving *n* variables is constructed from these *n* variables and logical operators such as NOT, AND, and OR. The SAT Problem is defined as the task of determining whether there exists an assignment *z* to the *n* variables such that the Boolean formula $\varphi(z)$ evaluates to true. A natural extension of SAT problem is the #SAT problem. Given a Boolean formula $\varphi$, the goal of the #SAT problem is to count the number of satisfying assignments for that formula. #SAT is a #P-complete problem that appears in various fields, including statistical estimation and statistical physics, particularly in contexts related to probability estimation. Since the SAT problem can be viewed as a special case of the #SAT problem (where the task is to check if there is at least one satisfying assignment), it is generally believed that the counting problem is more computationally challenging.

In a different aspect of complexity, Solomonoff[6,7,8,9], Kolmogorov[10,11], and Chaitin[12,13,14] defined the algorithmic complexity of a string as the length of the shortest binary program that can generate it. The precise definition is as follows:

Consider a universal computer *U*, the Kolmogorov complexity of a binary string *x* is defined as

$$K_U(x) = \min_{p:U(p)=x} l(p) \tag{1.1}$$

The minimum length over all programs that print *x* and halt. Thus, $K_U(x)$ is the shortest description length of *x* over all descriptions interpreted by computer *U*.

It has been demonstrated that the expected value of the Kolmogorov complexity of a random Bernoulli sequence is closely related to its Shannon entropy [15,16].

**Theorem 1.2** (Relationship between Kolmogorov complexity and entropy). Let $x_1, x_2, \ldots, x_{2^n}$ be drawn according to a Bernoulli$(\gamma)$ process. Then

$$\frac{1}{2^n} K(x_1, x_2, \ldots, x_{2^n}|2^n) \to H(\gamma) \text{ in probability.} \tag{1.2}$$

Kolmogorov made a crucial observation regarding the inherent independence of the complexity definition from specific computational models. The expected length of the shortest binary description for a random variable closely approximates its entropy. This leads to the conclusion that the shortest computer description serves as a universal coding mechanism, performing effectively across all probability distributions.

Kolmogorov complexity is a profound concept that provides key insights into computation and efficient algorithms [17,18]. It also offers a unique perspective on statistical inference and modeling. Solomonoff's inductive reasoning theory, a cornerstone of algorithmic information theory, proposes a method for predicting future data based on past observations by assigning probabilities to different hypotheses. Simpler hypotheses are given higher probabilities, as they can be generated by shorter programs. This combines Bayesian reasoning with algorithmic complexity, forming a powerful method for making predictions from incomplete data. Solomonoff's work has influenced many areas of machine learning and AI, especially in developing learning algorithms that generalize from limited data.

Out of the $2^{2^n}$ possible Boolean functions, the vast majority are complex and require exponentially sized circuits for computation. Since any SAT formula represents a Boolean function, any algorithm solving the SAT problem must first read the formula. If the formula is exponentially long, reading it alone will take exponential time, making it impossible for any algorithm to operate efficiently in polynomial time. In practice, we are generally only interested in functions described by concise (short, polynomial-length) SAT formulas. Theorem 1.1 indicates that such functions are extremely rare. This raises a natural question: do these special functions share common properties? Can these properties be further exploited? Although, To the best of our knowledge, previous literature has not explicitly addressed this question, we argue that it is crucial and holds significant theoretical and practical value. This article seeks to explore the SAT problem from this perspective, using Kolmogorov complexity theory as a bridge.

## 2. Two equivalent expressions of the information contained in Boolean function

The SAT problem is inherently connected to Kolmogorov complexity theory because its fundamental operations—AND, OR, and NOT—form a universal computation framework. Each SAT formula can be viewed as a universal computer or program that generates a bit string of a specified length $2^n$.

In accordance with Kolmogorov complexity theory, the information encapsulated within a specific SAT formula (corresponding to a Boolean function) can be articulated in two equivalent representations. One such representation is in the form of a program, exemplified by Program $\mathbb{p}$, which employs the loop syntax of the C programming language:

for（i= 0; i< $2^n$;i++） //Program $\mathbb{p}$
   {
       Given i as input, calculate the one-bit output of the SAT formula
       and print this bit.
   }

This program can output the $2^n$ bits corresponding to the function, generating a sequence of $2^n$ bits. The only two variables in this program are *n* and a specific short SAT formula expression, where the length of the expression is polynomial in nature. The overall length of this program can be expressed as:

$$l(p) = c + \log 2^n + l(\text{SAT formula}) = c + n + O(P(n)) = O(P(n)) \qquad (2.1)$$

Now we have a concise program of polynomial length that produces $2^n$ bits. According to Kolmogorov complexity theory, the shortest such program is inherently incomputable; it is impossible to determine definitively that this program is the shortest among all programs generating the same bit string. Instead, we can only establish an upper bound on the Kolmogorov complexity of that string.

$$K(x_1, x_2, \ldots, x_{2^n} | 2^n) \leq O(P(n)) \qquad (2.2)$$

At the same time, it was showed that the probability that a bit sequence can be compressed by more than *k* bits is no greater than $2^{-k}$ [15].

**Theorem 2.1** (Incompressible theorem). Let $x_1, x_2, \ldots, x_{2^n}$ be drawn according to a Bernoulli (1/2) process. Then

$$P(K(x_1, x_2, \ldots, x_{2^n} | 2^n) < 2^n - k) < 2^{-k} \qquad (2.3)$$

Therefore, the vast majority of sequences have a Kolmogorov complexity close to their length. Another way to express the information contained in a SAT formula is through the truth table of the corresponding Boolean function. Since the $2^n$ inputs in the truth table change systematically, incrementing from 0 to $2^n - 1$, the truth table can be simplified to the $2^n$ sequential output bits corresponding to these inputs. According to Theorem 2.1, the probability that a string can be compressed by more than $k$ bits does not exceed $2^{-k}$ of all possible $2^{2^n}$ output strings.

From the analysis presented above, a conclusion akin to Theorem 1.1 can be derived, among all possible $2^{2^n}$ functions (bit strings), the SAT formulas corresponding to the vast majority of functions must have a complexity close to their length of $2^n$. This is because bit strings that can be effectively compressed from exponential length to polynomial length are extremely rare. Therefore, Theorem 1.1 can be viewed as a special case of the incompressibility theorem in the circuit model.

Succinct SAT formulas with polynomial length must have corresponding output strings that can be effectively compressed. These strings can be compressed over a certain number of $2^n - O(P(n))$ bits. Given this context $O(P(n)) \ll 2^n$ when $n \to \infty$, we refer to them as effectively compressible.

The equivalence between Boolean formulas and circuit models is well-established in computational complexity theory. The class of effectively compressible languages under the Boolean formulas or circuit model is defined as follows:

**Definition 1.1** $C_P = \{x: \exists d, \mathbb{p}(d) = x, |d| = \text{poly}(n), |x| = 2^n\}$

In the expression above, $d$ represents the Boolean formulas or circuit description, $|d|$ denotes its length, and $\mathbb{p}(d)$ refers to the bit strings (Boolean functions) generated by Program $\mathbb{p}$ corresponding to the Boolean formulas.

The class of effectively compressible languages under Turing machines is strictly defined as follows:

**Definition 1.2** $K_P = \{x: \exists d, U(d) = x, |d| = \text{poly}(n), |x| = 2^n\}$

In the expression above, $d$ represents the Turing machine program, $|d|$ denotes the length of the program, and $U(d)$ represents the bit string generated by program $d$.

From the perspective of Kolmogorov complexity theory, a short SAT formula serves as an efficient compression program for a bit string of exponential length $2^n$. Both the SAT decision problem and the #SAT problem can be viewed as methods for distinguishing or determining different characteristic bit strings within an efficiently compressed representation of SAT formulas.

## 3. The Invariance Theorem and its potential natural extension

It is well established that the specific computational model is not crucial in studying computational complexity, as Turing machines can simulate all other models with at most polynomial slowdown [4]. A similar key theorem in Kolmogorov complexity is the invariance theorem, which is fundamental for the theory's development.

**Theorem 3.1** (Universality of Kolmogorov complexity). If $U$ is a universal computer, for any other computer $A$ there exists a constant $C_A$ such that

$$K_U(x) \leq K_A(x) + C_A \qquad (3.1)$$

for all strings $x \in \{0,1\}^*$, and the constant $C_A$ does not depend on $x$.

Broadly speaking, the invariance theorem indicates that all computational models capable of expressing partial recursive functions are approximately equally succinct. The remarkable utility and inherent validity of Kolmogorov complexity theory stem from this independence of the description method.

It is important to note that current definitions of Kolmogorov complexity are limited to the Turing machine model. Consequently, the invariance theorem is demonstrated through mutual simulations among different Turing machines, establishing invariance only within variations of the Turing model. In contrast, the circuit model has long been regarded as an equivalent model to Turing machines in the study of computational complexity [4]. The essence of the earlier program $\mathbb{p}$ is to utilize the Boolean circuit model to generate bit strings. The circuit complexity of a language is closely related to its time complexity; any language with small time complexity will also exhibit small circuit complexity, as illustrated by the following theorem.

**Theorem 3.2** Let $t: \mathbf{N} \to \mathbf{N}$ be a function, where $t(n) \geq n$. If $A \in \text{TIME}(t(n))$, then $A$ has circuit complexity $O(t^2(n))$.

Theorem 3.2 is fundamentally equivalent to Cook-Levin theorem.

**Theorem 3.3** (Cook-Levin theorem). SAT and Circuit-SAT are NP-complete.

A natural question arises: can Kolmogorov complexity be defined equivalently within the circuit model? Can the invariance theorem be extended to the circuit model? Since any program essentially expresses an algorithm, both Turing machines and SAT descriptions should effectively describe the same algorithm. Circuit complexity has been suggested as a prior in Solomonoff's theory of inductive inference [18]. Eric Allender [19] has made an 'outrageous statement that nonetheless contains a lot of truth':

"Kolmogorov Complexity is roughly the same thing as Circuit Complexity."

Therefore, this paper further conjectures that the two models of description are equivalent, namely:

**Proposition 3.1** $C_P = K_P$, For a given bit string $x$ of length $2^n$:

(1) $C_P \subseteq K_P$: If there exists a succinct SAT formula $\varphi$ that can output $x$ using Program $\mathbb{p}$, then $\varphi$ can be polynomial-time reduced to a succinct program that generates $x$.

(2) $K_P \subseteq C_P$: Conversely, if there exists a succinct program $P$ that produce $x$, it can be polynomial-time reduced to a succinct SAT formula corresponding to the same string.

Using the two complexity classes $C_P$ and $K_P$ defined earlier, Proposition 3.1 implies that $C_P = K_P$. The first part $C_P \subseteq K_P$ is clearly valid, as it can be reduced using Program $\mathbb{p}$. The key lies in the second part $K_P \subseteq C_P$. If Proposition 3.1 is proven false, there would be an essential difference between the circuit model and the Turing machine model. Although the final section of this paper explores this proposition, it does not provide a proof. The subsequent content is inspired by the assumption that this proposition holds, yet it also holds significance without it.

## 4. Basic distinguishability problem and limit of Solomonoff's inductive reasoning theory

Due to the interest in succinct SAT formulas, the corresponding bit strings of which can be effectively compressed, it is natural to examine which bit strings can be effectively compressed. Commonly, these bit strings can be roughly classified into two categories based on the ratio of the number of ones (or zeros) in the entire string. Let the bit string be $x = \{x_1, x_2, \ldots, x_{2^n}\}$ and $\sum_{i=1}^{2^n} x_i = k$, there are $k$ ones in the bit string.

**Type 1** (narrow periodicity). refers to bit strings that exhibit a specific repeating pattern. $k = O(P(n))$, ($k \ll 2^{n-1}$). According to Theorem 1.2, these strings can be effectively compressed

**Type 2** (general periodicity). refers to bit strings that, $k \approx 2^{n-1}$, while not strictly periodic, can still exhibit regular patterns and can be generated by short programs, such as: $\pi, e, \sqrt{2}$.

Under the distribution induced by the computer, the probability of obtaining a simple string is greater than the probability of obtaining a complex string of the same length. This observation leads to the concept of a universal probability distribution over bit strings [15]. The universal probability of the string $x$ is defined as:

$$P_U(x) = \sum_{p:U(p)=x} 2^{-l(p)} = \Pr(U(p) = x) \tag{4.1}$$

which is the probability that a program randomly drawn as a sequence of fair coin flips $p_1, p_2, \ldots$ will print out the string $x$.

Universal probability distributions encompass a mixture of all computable probability distributions. Notably, equation (4.1) shows that universal probability considers all programs that produce the same bit string as a whole or as an ensemble.

The concepts of universal probability and Kolmogorov complexity are equivalent, and their profound relationship is illustrated by the coding theorems [15,16].

**Theorem 4.1** (The Coding Theorem). There is a constant $c$ such that for every $x$,

$$-\log P_U(x) = K(x) \tag{4.2}$$

With equality up to an additive constant $c$.

In this way, Kolmogorov complexity and universal probability have the same status in the complexity as measure of general algorithms, and their relationship is simple:

$$P_U(x) \approx 2^{-K(x)} \tag{4.3}$$

This article further extends the concept of universal probability by considering all programs that produce sequences of $2^n$ bits with the same $k$ ones as an ensemble. This approach is motivated by two reasons: the bit strings share the same Kolmogorov complexity and universal probabilities, and if generated randomly, the programs do so with equal probability. More importantly, according to Theorem 1.2, these bit strings form Bernoulli distributions with the same parameter. A significant innovation in Kolmogorov complexity theory is that Turing machine programs alone are sufficient to define and quantify randomness, without the need for probability. This theory can be extended to define and measure probability distributions, especially the Bernoulli distribution. This justifies the proposal of both Boolean formula and Turing versions of the basic distinguishability problem. The second reason is discussed in Section 5, where a quantum version of the problem is introduced.

Let the bit string be $x = \{x_1, x_2, \ldots, x_{2^n}\}$ and $\sum_{i=1}^{2^n} x_i = k$, there are $k$ ones in the bit string. Define $\gamma = k/2^n$. Based on different values of $k$ in Type 1 language, the corresponding set of Boolean formulas and Turing programs can be further categorized into the following two sub-ensembles.

**Definition 4.1** $B(\gamma) = \{Boolean\ formulas\ d: \mathbb{p}(d) = x, |d| = poly(n), |x| = 2^n, \gamma = k/2^n\}$

**Definition 4.2** $P(\gamma) = \{Turing\ programs\ d: U(d) = x, |d| = poly(n), |x| = 2^n, \gamma = k/2^n\}$

$B(\gamma) \subseteq P(\gamma)$ holds because every Boolean formula in $B(\gamma)$ can be evaluated using program $\mathbb{p}$. $P(\gamma) \subseteq B(\gamma)$ will also hold if Proposition 3.1 holds. Definition 4.3 is a special case or subset of Definition 4.2, where we restrict our attention to the shortest Turing programs.

**Definition 4.3** $P_s(\gamma) = \{shortest\ Turing\ programs\ d_s: U(d_s) = x, |d_s| = poly(n), |x| = 2^n, \gamma = k/2^n\}$

These sub-ensembles can clearly be arranged in a sorted order as follows: $B(0)$, $B(1/2^n), B(2/2^n), B(3/2^n)\ldots$ and $P(0), P(1/2^n), P(2/2^n), P(3/2^n)\ldots$ To represent the order, we introduce the ">" symbol. We say that $B(\gamma_1) > B(\gamma_0)$ if $(\gamma_1 > \gamma_0)$. Based on this, the following basic distinguishability problem is defined from the perspective of Boolean formula ensembles.

**Definition 4.4** (Basic distinguishability(BD) Problem in the Boolean formula). The BD problem in the Boolean formula version involves distinguishing between two Bernoulli distributions: Bernoulli$(\gamma_0)$, induced by the $B(\gamma_0)$ ensemble, and Bernoulli$(\gamma_1)$, induced by the $B(\gamma_1)$ ensemble.

The Boolean formula BD problem is based on the problem of determining the proportion of 1's in the output bitstring generated by a given Boolean formula.

**Definition 4.5** (The Program Output Statistical Prediction (POSP) problem, in Boolean formula version) Given a Boolean formula $d$, let $\mathbb{p}(d) = x$, where $|x| = 2^n$. The problem is to determine whether the proportion of 1's in $x$ is greater than a given threshold γ, where $0 < \gamma < 1$。

This article argues that the POSP problem holds significant research value, particularly because it can be directly related to certain complexity classes in computational complexity. The following proposition is presented.

**Proposition 4.1** The Boolean formula version of the POSP problem is #P complete.

**Proof:** It is necessary to prove that these two problems can mutually reduce to each other. The #SAT problem is #P-complete, binary search algorithm is used to reduce the Boolean formula version POSP problem to the #SAT problem. Suppose there exists an algorithm **A** for the POSP problem. The reduction from algorithm **A** to an algorithm **B** for the #SAT problem can be designed as follows: we first compare Bernoulli(0.5) using algorithm **A**. If it is greater than Bernoulli(0.5), we then compare Bernoulli ((0.5+1)/2=0.75); if it is less, we compare Bernoulli((0.75+0.5)/2=0.625). This process continues until the binary search converges on Bernoulli($(k/2^n)$). The time complexity of this reduction, based on the binary search algorithm, is polynomial.

On the other hand, the #SAT problem can be obviously reduced to the POSP problem. □

In the same time, we propose the following Proposition.

**Proposition 4.2** The Boolean formula version POSP problem can be reduced to the BD problem.

**Proof:** Suppose there exists an algorithm **A** that solves the POSP problem. We can design a reduction from algorithm **A** to an algorithm **B** for the BD problem as follows: first use algorithm **A** to compare with Bernoulli(0.5). If both are greater than Bernoulli(0.5), compare with Bernoulli ((0.5+1)/2=0.75); if both are less, compare with Bernoulli((0.75+0.5)/2=0.625). This binary search continues until we find a value such that one is greater and the other is smaller. □

The SAT problem can be viewed as a simplified version of the BD problem, requiring only a single invocation of the algorithm for POSP problem. The ensemble of unsatisfiable SAT formulas corresponds to Bernoulli(0) distribution, while the ensemble of satisfiable SAT formulas corresponds to Bernoulli($0 < \gamma \leq 1$) distribution, with these distributions induced by the SAT formula ensemble.

If Proposition 3.1 holds, which states that SAT formulas and their short program descriptions can be reduced to each other in polynomial time, then the Turing machine version of the basic distinguishability problem is equivalent to the SAT formula version of the BD problem.

**Definition 4.6** (BD Problem in the Turing Machine version). This problem involves distinguishing between two Bernoulli distributions: Bernoulli($\gamma_0$), which is induced by the $P(\gamma_0)$ ensemble, and Bernoulli($\gamma_1$) which is induced by the $P(\gamma_1)$ ensemble.

**Definition 4.7** (POSP problem in the Turing Machine version) given a program $d$, let $U(d) = x$, and $|x| = 2^n$. The problem of determining the proportion of 1's in $x$ is greater than a threshold $\gamma$, where $0 < \gamma < 1$。

This paper argues that the program version of the BD problem is a fundamental issue in mathematics, regardless of its relation to the SAT problem. Its inherent importance makes it a critical area of study. It raises the question of whether efficient algorithms exist within the program ensemble space, especially since no efficient distinguishability algorithm is evident in the probability space.

For type 1 bit strings, universal probabilities can be used to define the time complexity of any algorithm that addresses the specific program version of the basic distinguishability problem: distinguishing between a Bernoulli(0) distribution and other Bernoulli distributions generated by program ensemble $P(\gamma)$.

$$\sum_k C_{2^n}^k . P_U(E_k) . C(E_k), \quad k=2^n . \gamma \tag{4.7}$$

In this context, we define the sequence ensemble $E_k$ as the sequence containing $k$ ones, which is generated by the program ensemble $P(\gamma)$. The associated time complexity $C(E_k)$ refers to the resources required to distinguish the ensemble $P(\gamma)$ from a reference ensemble $P(0)$, where the corresponding sequence of $P(0)$ consisting entirely of zeros. $C_{2^n}^k$ represents the number of combinations for $E_k$. It is worth noting that there exists a significant theorem regarding the average-case complexity of universal distributions in Kolmogorov complexity theory [20].

**Theorem 4.2** (universal distribution average complexity). Let **A** be an algorithm with inputs in $\{0,1\}^\infty$. If the inputs to **A** obeys the universal distribution $P_U$, then the average-case time complexity of **A** is of the same order of magnitude as the corresponding worst-case time complexity.

By applying this theorem to the program version of the BD problem, we can draw the following conclusion:

**Proposition 4.3** For any algorithm solving the program version distinguishability problem, the average-case time complexity is of the same order as the worst-case complexity.

**Proof:** In the program version of the basic distinguishability problem, the inputs follow the universal distribution $P_U$. Therefore, for any algorithm addressing this problem, the average complexity is equivalent to the worst-case complexity. □

Regarding the Turing Machine version BD problem, any algorithm must be capable of distinguishing among all possible input programs. However, these programs have different probabilities of occurrence based on their length. As shown in Equation (4.3), the probability of encountering the shortest program dominates within the program ensemble. Specifically, the likelihood of selecting the shortest program among all input programs that produce a given sequence approaches 1. The average complexity of any algorithm corresponds to the shortest input program, as established by the universal distribution average complexity theorem. Thus, the analysis of the Turing Machine version BD problem can be simplified to the shortest program version of the BD problem. If a hypothetical arbitrary algorithm can solve the program version of the BD problem, it should also be able to solve the shortest program version.

**Definition 4.8** (BD Problem in the shortest program version). This problem involves distinguishing between two Bernoulli distributions: Bernoulli($\gamma_0$), which is induced by the $P_s(\gamma_0)$ ensemble, and Bernoulli($\gamma_1$) which is induced by the $P_s(\gamma_1)$ ensemble.

By simplifying the problem in this manner, the complexity analysis is expected to become more straightforward. Any shortest program exhibits two important properties, one of which is [12]:

**Theorem 4.3** Any shortest programs are inherently random (incompressible) strings.

Another property of the shortest programs represents the second cornerstone—more accurately, an obstacle—of Kolmogorov complexity theory: the incomputability theorem [16]

**Theorem 4.4** (Incomputability of Kolmogorov Complexity). The function $K(x)$ is not recursive. Moreover, no partial recursive function $\phi(x)$ defined on an infinite set of points can coincide with $K(x)$ over the whole of its domain of definition.

Solomonoff suggests that observed data and hypotheses can be represented as bit strings, with each hypothesis corresponding to a computer program that generates sequences matching the observed data. Any scientific theory involves two steps: generating alternative hypotheses based on observations and selecting the most likely one. Statistics provides the mathematical foundation for this process. Solomonoff's theory of inductive reasoning helps make predictions by considering all possible hypotheses and their complexities. Inductive reasoning theories boil down to two key elements: Bayes' law and universal prior probability.

In this framework, a hypothesis can be viewed as a probability distribution (measure) $\mu$ over a sample space of infinite sequences. This distribution $\mu$ represents the concept or phenomenon being investigated, and the goal is to predict outcomes related to this phenomenon $\mu$. We assume a prior distribution $\mu$ over the sample space $S$, where $\mu(x)$ denotes the probability of a sequence starting with $x$. The final probability $\mu(y|x)$ represents the probability of the next symbol or string being $y$, given the initial string $x$. This relationship can be expressed using Bayes' law as follows:

$$\mu(y|x) = \frac{\mu(xy)}{\mu(x)} \qquad (4.8)$$

Here, $\mu(xy)$ is the joint probability of the sequence starting with $x$ and followed by $y$, while $\mu(x)$ is the marginal probability of the sequence starting with $x$. This framework allows us to update our predictions based on new information effectively. It turned out that $P_U(y|x)$ is the perfect estimator for any computable $\mu(y|x)$. The expected values of the sum of the differences between any computable hypothesis and estimated probabilities are bounded by a constant, which is a very powerful result that Solomonoff referred to as the Convergence Theorem [7,8,16].

**Definition 4.9** Let $B$ be a finite alphabet, and let $x$ be a word over $B$. The summed expected squared error at the $n$th prediction $S_n$ is defined by

$$S_n(a) = \sum_{l(x)=n-1} \mu(x)\left(\sqrt{P_U(a|x)} - \sqrt{\mu(a|x)}\right)^2 \quad (4.9)$$

$$S_n = \sum_{a \in B} S_n(a) \quad (4.10)$$

Let $K(\mu)$ be the length of the shortest program computing the function $\mu$.

**Theorem 4.5** (the Convergence Theorem) Let $\mu$ be a computable measure. Then, $\sum_n S_n < k$ with constant $k = K(\mu)\ln 2$.

Theorem 4.5 showed that conditional prior probability $P_U(y|x)$ suffices to approximate any computable distribution $\mu(y|x)$ and the convergence is very fast [16]. Ultimate compression and the shortest programs have been shown to almost always lead to optimal prediction.

Solomonoff's inductive formula $P_U(y|x)$ can be formally expressed as follows:

$$P_U(y|x) = \frac{P_U(xy)}{P_U(x)} \quad (4.11)$$

The problem with Bayes' rule has always been the determination of the prior. Using $P_U$ universally gets rid of that problem and is provably perfect. To estimate the actual probabilities $\mu(y|x)$ to predict outcomes $y$ given a sequence of observed outcomes $x$, can be viewed as a mathematical embodiment of Occam's Razor, which serves as a fundamental guiding principle in scientific research [15,16].

"Find all the rules fitting the data and then predict $y$ according to the universal distribution on them."

Theories, hypotheses, and algorithms must be developed to make predictions based on an initial sequence of data. Solomonoff's universal prediction theory suggests that the universal distribution is well-suited for making these predictions. However, the shortest programs correspond to incompressible (random) strings, which lack regularity or patterns that can be exploited. As a result, it becomes impossible to design algorithms that can take advantage of patterns or rules that do not exist. In fact, the incompressibility of random objects yields a simple but powerful proof technique [16].

The universal distribution of the shortest program is itself, which leads to self-reference when using Solomonoff's inductive reasoning to predict the output of the shortest program. In other words, to determine the output of the shortest program, we are effectively referencing the program itself, creating a circular dependency. From the properties of the shortest programs and the coding theorem, we can draw an important conclusions that highlight the natural limitations of Solomonoff's inductive reasoning theory.

**Proposition 4.4:** (The limit of program output prediction (POP) problem) Essentially, the only "algorithm" capable of predicting the output $x$ of shortest program $p_x$ is the shortest program itself. Any other algorithms $A$ will be computationally indistinguishable from the universal computer $U$.

**Proof:** Suppose for the sake of contradiction that there exists an algorithm $A$ that satisfies:

For any output $x$, $A(p_x) = U(p_x)$, where $p_x$ is the minimal program for $x$.

For non-minimal programs $q \neq p_x$, $A(q)$ may output arbitrarily incorrect results.

Define $D_A$ as the output distribution of $A$ under the universal distribution $P_U$, and $D_U$ as the output distribution of $U$.

By the Coding Theorem:

$$P_U(x) = \sum_{p:U(p)=x} 2^{-l(p)} \geq 2^{-K(x)+O(1)} \quad (4.12)$$

where $\ell(p)$ denotes the length of program $p$. and $K(x)$ is the Kolmogorov complexity of $x$. For minimal programs $p_x$, $l(p_x) = K(x)$, so the contribution of minimal programs to $D_U$ is dominated by the term $2^{-K(x)}$.

Thus, the distribution $D_U$ can be written as:

$$D_U(x) = P_U(x) = 2^{-K(x)+O(1)} + \varepsilon_x \quad (4.13)$$

where $\varepsilon_x$ accounts for non-minimal programs, and satisfies $\varepsilon_x \leq 2^{-K(x)-O(1)}$.

Since $A$ only predicts $U(p_x)$ correctly, and behaves arbitrarily otherwise, the distribution $D_A(x)$ is:

$$D_A(x) = 2^{-K(x)+O(1)} + \delta_x \quad (4.14)$$

where $\delta_x$ captures errors from non-minimal programs. By assumption, $\delta_x$ is negligible under $P_U$, as non-minimal programs contribute insignificantly to $P_U$.

Now consider any Probabilistic Polynomial Time (PPT) distinguisher $T$, the advantage of T distinguishing $D_A$ from $D_U$ satisfies:

$$|P_r[T(D_A) = 1] - P_r[T(D_U) = 1]| \leq \sum_x |D_A(x) - D_U(x)| \cdot Adv_D(x) \quad (4.15)$$

where $Adv_D(x)$ is the distinguisher's advantage on input $x$.

Next, we bound the statistical distance between $D_A$ and $D_U$:

$$\Delta(D_A, D_U) = \sum_x |D_A(x) - D_U(x)| \leq \sum_x (\varepsilon_x + \delta_x). \quad (4.16)$$

From the previous arguments, we know that $\varepsilon_x + \delta_x \leq 2^{-K(x)-O(1)}$, which decays exponentially with $K(x)$.

Summing over all $x$, we get:

$$\sum_x 2^{-K(x)-O(1)} \leq 2^{O(1)} \sum_x 2^{-K(x)} = 2^{-O(1)} \quad (4.17)$$

This follows from the fact that $\sum_x 2^{-K(x)} \leq 1$, which is a consequence of Kraft's inequality and the Prefix Turing Machine.

Therefore, the total advantage of the distinguisher is negligible, and no PPT distinguisher $T$ can achieve non-negligible advantage. This implies that $D_A$ and $D_U$ are computationally indistinguishable under $P_U$.

Thus, it follows that the only algorithm capable of predicting the output $x$ of the shortest program $p_x$ is the shortest program itself. No algorithm other than the shortest program $p_x$ can distinguishably predict the output. Therefore, algorithm $A$ cannot exist unless it is functionally equivalent to $U$. □

Although the above proof is formulated for the entire set of $x$, the conclusion of the above proof holds for any subset $x$ (e.g., the subset of all-zero sequences) as well.

Solomonoff's inductive reasoning theory, Gödel's incompleteness theorem, and the halting problem all involve forms of self-reference, though they stem from different contexts and lead to distinct logical or computational challenges. Induction and deduction are two fundamental categories of reasoning, with understanding inductive reasoning being a problem that has engaged mankind for thousands of years. This issue has attracted significant attention across a wide range of fields and is central to the philosophy of science. Solomonoff Induction offers a formal framework that integrates algorithmic information theory and Bayesian methods to address this complexity [21]. While the Gödel incompleteness theorem is often seen as revealing a limitation of deductive logic, Proposition 4.4 might be viewed as a potential fundamental limitation of inductive reasoning.

In the context of predicting specific statistical properties of an output sequence, determining or predicting whether a program produces $2^n$ zeros is equivalent to checking if it follows a Bernoulli(0) distribution. Consequently, applying Proposition 4.4 to this problem leads us to conclude that there exists a unique non-effective algorithm for checking whether a program outputs $2^n$ zeros. This conclusion can be further inferred to yield the following proposition.

**Proposition 4.5** There exists no algorithm with polynomial time complexity for solving the POSP problem related to the shortest programs .

**Proof:** Assuming there exists a polynomial-time algorithm **A** that solves this problem, we can reduce algorithm **A** to an effective decision procedure **B** that determines whether the program outputs $2^n$ zeros, using a method similar to that in Proposition 4.1. This would create a contradiction with Proposition 4.4. □

**5. The quantum version of the basic distinguishability problem**

The primary objective of Bayesian theory is to develop reliable reasoning and decision-making methods in the face of incomplete information. Proposition 4.4 highlights a crucial limitation of Bayes' rule. Interestingly, there are parallels between Bayesian reasoning and quantum mechanics [22]. While classical probability theory relies on the 1-norm, quantum mechanics can be seen as an extension of probability theory that operates under the 2-norm framework [23]. Furthermore, nonlinear quantum mechanics suggests that polynomial-time solutions may exist for NP-complete and #P problems [24]. Although the quantum version of the basic distinguishability problem is not explicitly framed in computational terms, it has been thoroughly investigated within the field of quantum physics.

The quantum basic distinguishability problem has been rigorously examined through the lens of statistical distance [25]. John Archibald Wheeler [26] has highly praised the contributions of this article, stating, "No one has done more than William Wootters to open up a pathway from information to quantum theory. " Wootters' work, being both accessible and directly related to this problem, serves as a valuable reference. We will leverage his findings while reframing them from the perspective of quantum computation.

When tossing a coin 100 times and getting "heads" 30 times, one might conclude that the probability of heads is about 0.30, suggesting the coin may be biased. However, due to statistical fluctuations in a finite sample, the true probability could actually be closer to 0.26 or 0.34.This situation mirrors quantum measurements. When analyzing a finite set of identically prepared quantum systems using a fixed measurement device, the observed frequencies of outcomes may differ from the actual probabilities due to statistical error. Consequently, it can be challenging to differentiate between two slightly different preparations of the same quantum system within a limited number of trials. We can consider two preparations indistinguishable if the difference in their actual probabilities is smaller than the typical statistical fluctuation.

**Definition 5.1** (The quantum version of the basic distinguishability problem). Distinguish two different Bernoulli distributions that are induced by two preparations of quantum system ensembles.

Imagine a beam of photons prepared by a polarizing filter and analyzed by a nicol prism. Let $\theta \in [0, \pi]$ be the angel by which the filter has been rotated around the axis of the beam. Each photon, when it encounters the nicol prism, has exactly two options: to pass straight through the prism (call this "yes" outcome) or to be deflected in a specific direction characteristic of the prism (the "no" outcome). By counting how many photons yield each of the two possible outcomes, an experimenter can learn something about the value of $\theta$ via the formula.

$$p = \cos^2 \theta \tag{5.1}$$

where $p$ is the probability of "yes".

Then, because of the statistical fluctuations associated with a finite sample ($m$), the observed frequency of occurrence of "yes" is only an approximation to the actual probability of "yes". More precisely, the experimenter's uncertainty in the value of $p$ is

$$\Delta p = \left[\frac{p(1-p)}{m}\right]^{1/2} \tag{5.2}$$

This uncertainty causes the experimenter to be uncertain of the value of $\theta$ by an amount.

$$\Delta \theta = \left|\frac{dp}{d\theta}\right|^{-1} \Delta p = \left|\frac{dp}{d\theta}\right|^{-1} \left[\frac{p(1-p)}{m}\right]^{1/2} \tag{5.3}$$

Thus each value of $\theta$ can be associated with a region of uncertainty, extending from $\theta - \Delta \theta$ to $\theta + \Delta \theta$. Let us call two neighboring orientations $\theta$ and $\theta'$ distinguishable in $m$ trials if their regions of uncertainty do not overlap, that is, if

$$|\theta - \theta'| \geq \Delta \theta + \Delta \theta' \tag{5.4}$$

This idea of distinguishability is then used to define a notion of distance, called "statistical distance" between quantum preparation. The statistical distance $d(\theta_1, \theta_2)$ between any two orientations $\theta_1$ and $\theta_2$ is defined to be

$$d(\theta_1, \theta_2) = \lim_{m \to \infty} \frac{1}{\sqrt{m}} \times [\text{maximum number of intermediate orientations} \\ \text{each of which is distinguishable (in } m \text{ trials) from its neighbors}] \tag{5.5}$$

This statistical distance is obtained essentially by counting the number of distinguishable orientations between $\theta_1$ and $\theta_2$.

From Eqs. (5.2)-(5.5) we obtain the following expression for statistical distance in terms of the function $p(\theta)$

$$d(\theta_1, \theta_2) = \frac{1}{\sqrt{m}} \int_{\theta_1}^{\theta_2} \frac{d\theta}{2\Delta\theta} = \int_{\theta_1}^{\theta_2} d\theta \frac{|dp/d\theta|}{2[p(1-p)]^{1/2}} \tag{5.6}$$

Upon substituting the actual form of the probability law $p(\theta) = \cos^2 \theta$ into this expression, the statistical distance become

$$d(\theta_1, \theta_2) = \theta_2 - \theta_1 \tag{5.7}$$

In fact, the concept of statistical distance can be defined on any probability space and is quite independent of quantum mechanics [24]. This serves as the second reason for proposing the SAT version and Turing version of the basic distinguishability problem, wherein other computational models replace the quantum mechanism as discussed earlier.

In a case where there are exactly two possible outcomes, the probability space is one-dimensional, every coin being characterized by its probability of heads. The statistical distance $d(p_1, p_2)$ between two coins with probability $p_1$ and $p_2$ of heads is defined in the following way:

$$d(p_1,p_2) = \lim_{n\to\infty} \frac{1}{\sqrt{m}} \times [\text{maximum number of mutually distinguishable} \quad (\text{in } m \text{ trials}) \text{ intermediate probabilities}] \tag{5.8}$$

Here two probabilities $p$ and $p'$ of heads are called distinguishable in $m$ trials if
$$|p - p'| \geq \Delta p + \Delta p' \tag{5.9}$$
where, as before
$$\Delta p = \left[\frac{p(1-p)}{m}\right]^{1/2} \tag{5.10}$$
then the following expression for statistical distance can be obtained:

$$d(p_1,p_2) = \int_{p_1}^{p_2} \frac{dp}{2[p(1-p)]^{1/2}} = \cos^{-1}(p_1^{1/2} p_2^{1/2} + q_1^{1/2} q_2^{1/2}) \tag{5.11}$$

where $q_1 = 1 - p_1$ and $q_2 = 1 - p_2$.

One can obtain the statistical distance by counting the number of these curves that "fit" between two given points in probability space. In a sense this is the most natural notion of distance on probability space, since it takes into account the actual difficulty of distinguish different probabilistic experiments. in the same time, the statistical distance between $p_1$ and $p_2$ is the shortest distance along the unit sphere. This shortest distance is equal to the angel between the unit vectors. The statistical distance is the "entropy differential metric" introduced by Rao in 1945 [15], the distance between two distributions arising from the Riemannian metric over the parameter space with the Fisher information metric tensor.

The main result of [25] is that the (absolute) statistical distance between two preparations is equal to the angel in Hilbert space between the corresponding rays.

**Theorem 5.1**: The statistical distance in Hilbert-space and probability space are equivalent and invariant under all unitary transformations.

The angel in Hilbert space is the only Riemannian metric on the set of rays, up to a constant factor, which is invariant under all unitary transformations (computed by any quantum computer), that is, under all possible time evolutions. In this sense, it is a natural metric on the set of states. It's as if nature defines distance between states by counting the number of distinguishable intermediate states [25].

As we know, any quantum computation is composed of three basic steps [27]: preparation of the input state, implementation of the desired unitary transformation acting on the state and measure of the output state.

If a quantum computer is utilized to solve the quantum version of the basic distinguishability problem, Theorem 5.1 leads us to an important conclusion that corresponds to a fundamental limit of quantum mechanics. Specifically, distinguishing between two preparations in Hilbert space is equivalent to distinguishing them in probability space. This is because the angle in Hilbert space remains invariant under all unitary transformations. Then measurement is the only quantum "algorithm" that remains. How many measurements are required to reliably distinguish between the two quantum states under a given measurement strategy?

Suppose $p_1 = 0$, we have

$$\Delta p_1 = \left[\frac{p_1(1-p_1)}{m}\right]^{1/2} = 0; \quad \Delta p_2 = \left[\frac{p_2(1-p_2)}{m}\right]^{1/2} \tag{5.12}$$

in order to distinguish them, we have

$$|p_2 - p_1| \geq \Delta p_1 + \Delta p_2 \Rightarrow$$
$$p_2 \geq \left[\frac{p_2(1-p_2)}{m}\right]^{1/2} \Rightarrow m \geq \frac{1-p_2}{p_2} \tag{5.13}$$
$$\Rightarrow if\ p_2 = 2^{-n}\ then\ m \geq 2^n - 1$$

In the worst situation, the expected trails in order to distinguish two preparations will increase exponentially with $n$. In the case of quantum computation, this means if two preparations are close enough, for example $|p_2 - p_1| = 2^{-n}$, the measurement must be performed exponential times in order to distinguish them. In this way, the quantum version basic problem cannot be efficiently solved by any quantum computer.

This indistinguishability (with one measurement) of non-orthogonal quantum states is at the heart of quantum computation and quantum information. It is the essence of the assertion that a quantum state contains hidden information that is not accessible to measurement, and thus plays a key role in quantum algorithms and quantum cryptography [27].

## 6. Equivalence of Kolmogorov Complexity in Turing Machines and Circuit Models

Proposition 3.1 questions the Kolmogorov equivalence between the circuit model and the Turing machine model. If this equivalence does not hold, the implications would still be profound, suggesting that the circuit model is fundamentally different from the Turing machine.

The first part of Proposition 3.1 is straightforward: the SAT formula functions as a low-level program, making it easy to simulate a SAT formula using a Turing machine. The key focus is the second part: if there exists a succinct program that generates a specific $2^n$ bit string, then we can create a corresponding succinct CNF formula that produces the same bit string (using Program $\mathbb{p}$) through polynomial-time reduction.

To design this polynomial-time reduction, we lack detailed information about the succinct program. Therefore, the only approach available is to adopt the technique used in the Cook-Levin theorem, which involves reductions based on computation history. The main proof idea of the Cook-Levin theorem is to simulate a Turing machine using a circuit model. To derive a smaller formula, the proof relies on two facts:
(1) The Turing machine *M* runs in polynomial time;
(2) Each basic step of a Turing machine is highly local, meaning that it examines and

changes only a few bits on the machine's tapes. The correctness of these local steps is expressed using smaller Boolean formulas.

Among the two facts used in the Cook-Levin Theorem, in the case of the expected proof, Fact 2 (that Turing machines are highly local) still holds, but Fact 1 does not apply. To generate an exponential-length bit string, the time complexity of the Turing machine must be exponential as well. Therefore, new facts need to be identified that can help construct smaller formulas. It has been shown that the local checkability of computation will be key in resolving the P versus NP problem [4].

It is important to note that the core part of Program $\mathbb{p}$ is a loop that runs for an exponential number of iterations. Regarding a succinct Turing machine program that generates a specific exponential-length bit string, it can be demonstrated that there exists a loop running for an exponential number of iterations within the short program.

**Proposition 6.1:** A succinct program that outputs a specific $2^n$ bit string must contain a loop that runs for an exponential number of iterations.

Before proving this proposition, we will first review the concept of a loop. Any loop structure consists of three essential elements: a counter variable, the loop body, and a test condition. This structure can be transformed into the following standard form in the C programming language.

```
    i=n1;
    while(i<n2)
    {
        The loop body;// function f(i)
        i=i+1;
    }
```

In the above program, $i$ is the counter variable, which is initialized with a value $n1$. The condition "while ($i < n2$)" serves as the test condition for the loop. To ensure that this loop runs for an exponential number of iterations, the difference $n2 - n1$ must be of exponential order.

Regarding the loop body, there is another important fact: the time complexity of executing the loop body once is polynomial. This restriction is quite natural, as the loop corresponds to short periodicity(regularity) in the string. Consequently, the loop body is used to express this short periodicity.

**Proof:** Suppose the length $l$ of the succinct program is polynomial $O(P(n)) \ll 2^n$. Since a Turing machine is equivalent to any modern computer language, and all modern languages consist of only three basic structures—sequential, branching, and looping—these structures can express any algorithm. If this program only consists of sequential and branching structure, it'll output at most $l \ll 2^n$ bits. Therefore it must contain loop structure. Since it can output at most $l$ bits in one loop iteration, the number of times the loop executes must be at least $2^n/l$, which is exponential in nature. In this analysis, we considered only one loop; however, the result remains unchanged even if there are additional loops.

As supporting evidence for this proof, the core of the currently known algorithms for $\pi$ and $e$ are all based on iterative structures. The essential function of any loop structure is to enable the reuse of specific code or rules within the loop body. Compared to the local checkability involved in the Cook-Levin theorem, loops represent another important form of local checkability, which is considered crucial for resolving the P versus NP problem [4]. Among the three basic programming structures, only loop structures can effectively reduce the overall length of a program. When the core of a certain algorithm is based on a loop structure, the resulting bit string must exhibit a certain regularity.

Incompressible sequences are random and do not possess any regularities or patterns; otherwise, we could reduce their description length by leveraging these regularities. Conversely, compressible sequences must demonstrate specific patterns or regularities. So, what exactly are these regularities? As Kolmogorov pointed out [10]:

"It is natural to recall that the absence of periodicity is, according to common sense, a symptom of randomness."

In Kolmogorov's work [9], his primary concern was on defining and measuring randomness. While he used the term "periodicity", he did not offer a formal definition for this concept. This absence of definition creates a gap in understanding how periodicity is negatively correlated with randomness. From our perspective, "periodicity" stands in contrast to randomness; it represents regularity. However, this notion of periodicity extends beyond the following simple mathematical definitions typically associated with it.

A periodic function $y = f(x)$, having a period $P$, can be represented as $f(x + P) = f(x)$.

By interpreting Kolmogorov's conclusions from an alternative perspective, we observe that sequences characterized by inherent regularities exhibit a certain degree of periodicity. Algorithms can be developed to harness this periodicity for the more efficient description of these sequences. In this context, periodicity is equivalent to regularity, which can be represented by loop structures that effectively minimize the description length of the sequences.

Therefore, we can define general periodicity as follows: The repeatable rules that can be described by a cyclic program. This general periodicity includes type 1 narrow periodicity. According to Proposition 6.1, the core components of short programs are loops. However, this loop is not necessarily the same as the loop in Program ℙ. While how to utilize this property to extend the proof of the Cook-Levin theorem for obtaining short SAT formulas remains to be explored, from our perspective, it represents a potentially promising avenue for future research.

## 7. Conclusion

This article is primarily based on two key concepts: traditional randomness is measured through probability, while Kolmogorov complexity theory offers an alternative measure based on universal computation that does not rely on probability distributions. These two measures of randomness have been theoretically shown to be equivalent. Consequently, this study aims to extend the concept of the shortest program for a universal computer to evaluate the basic Bernoulli distribution. Additionally, there is a clear, natural, and profound connection between the SAT problem and Kolmogorov complexity theory, which leads us to propose a concise research idea.

1. A SAT formula can be conceptualized as a low-level program constructed from

basic logic gates, with each formula representing a specific Boolean function. A concise SAT formula can generate a bit string that is amenable to effective compression, thereby establishing a significant connection between the SAT problem and Kolmogorov complexity theory. This suggests that the entities analyzed in both the SAT problem and Kolmogorov complexity theory are fundamentally equivalent.

2. By examining which bit strings can be efficiently compressed and represented as concise SAT expressions, we discover that the #SAT problem creates a fundamental distinguishability challenge: differentiating between the Bernoulli distributions induced by ensembles of SAT expressions. From this viewpoint, three versions of the basic distinguishability problem are proposed. Notably, the shortest program version of this problem can be shown to have only one "algorithm".

3. The quantum version of the basic distinguishability problem involves Bernoulli distributions induced by quantum mechanisms. This can be determined using the equivalence of statistical distances in Hilbert space and information space. Since this distance is invariant under any unitary transformation, we can establish a lower bound on the computational complexity of this problem.

4. The Cook-Levin theorem demonstrates the computational complexity equivalence of Turing machines and circuit models through reductions based on computation history. This suggests that it should be theoretically possible to extend this concept to establish the Kolmogorov complexity equivalence between the two models.

REFERENCE

[1] Stephen A. Cook. 1971. The complexity of theorem-proving procedures. In Proceedings of the third annual ACM symposium on Theory of computing - STOC '71. DOI:https://doi.org/10.1145/800157.805047

[2] B.A. Trakhtenbrot. 1984. A Survey of Russian Approaches to Perebor (Brute-Force Searches) Algorithms. IEEE Annals of the History of Computing (October 1984), 384–400. DOI:https://doi.org/10.1109/mahc.1984.10036

[3] Richard M. Karp. 1972. Reducibility among Combinatorial Problems. In Complexity of Computer Computations. 85–103. DOI:https://doi.org/10.1007/978-1-4684-2001-2_9

[4]Sanjeev Arora and Boaz Barak. 2009. Computational Complexity: A Modern Approach. Cambridge University Press eBooks,Cambridge University Press eBooks. DOI:https://doi.org/10.1017/cbo9780511804090

[5] Lance Fortnow. 2009. The status of the P versus NP problem. Communications of the ACM (September 2009), 78–86. DOI:https://doi.org/10.1145/1562164.1562186

[6] RayJ. Solomonoff. 2001. A PRELIMINARY REPORT ON A GENERAL THEORY OF INDUCTIVE INFERENCE. (January 2001).

[7] R. Solomonoff. 1978. Complexity-based induction systems: Comparisons and convergence theorems. IEEE Transactions on Information Theory (July 1978), 422–432. DOI:https://doi.org/10.1109/tit.1978.1055913

[8] R. J. Solomonoff. 2003. The Kolmogorov Lecture The Universal Distribution and Machine Learning. The Computer Journal 46, 6 (June 2003), 598–601. DOI:https://doi.org/10.1093/comjnl/46.6.598

[9] R.J. Solomonoff. A formal theory of inductive inference. Part II. Information and Control 7, 2 , 224–254. DOI:https://doi.org/10.1016/s0019-9958(64)90131-7

[10] A. Kolmogorov. Logical basis for information theory and probability theory. IEEE Transactions on Information Theory, 662–664. DOI:https://doi.org/10.1109/tit.1968.1054210

[11]A. N. Kolmogorov 1968. Three approaches to the quantitative definition of information , International Journal of Computer Mathematics, 2:1-4, 157-168, DOI:https://doi.org/10.1080/00207166808803030

[12] Gregory J. Chaitin 1966. On the Length of Programs for Computing Finite Binary Sequences. Journal of the ACM, 547–569. DOI:https://doi.org/10.1145/321356.321363

[13] Gregory J. Chaitin. 1974. Information-Theoretic Limitations of Formal Systems. Journal of the ACM (July 1974), 403–424. DOI:https://doi.org/10.1145/321832.321839

[14] Gregory J. Chaitin. 1975. Randomness and Mathematical Proof. Scientific American (May 1975), 47–52. DOI:https://doi.org/10.1038/scientificamerican0575-47

[15] Thomas M. Cover and Joy A. Thomas. 1991. Elements of Information Theory. Wiley Series in Telecommunications. DOI:https://doi.org/10.1016/0165-1889(95)00877-2

[16] Ming Li and PaulM.B. Vitányi. 1993. An introduction to Kolmogorov complexity and its applications. Texts and monographs in computer science,Texts and monographs in computer science (January 1993).

[17] Lance Fortnow. 2004. Kolmogorov Complexity and Computational Complexity. (January 2004).

[18] Cole Wyeth and Carl Sturtivant. 2023. A Circuit Complexity Formulation of Algorithmic Information Theory. Physica D: Nonlinear Phenomena,Volume 456 (June 2023), https://doi.org/10.1016/j.physd.2023.133925.

[19] Eric Allender. 2008. Circuit Complexity, Kolmogorov Complexity, and Prospects for Lower Bounds. (January 2008).

[20] Ming Li and Paul M.B. Vitányi. 1992. Average case complexity under the universal distribution equals worst-case complexity. Information Processing Letters 42, 3 (May 1992), 145–149. DOI:https://doi.org/10.1016/0020-0190(92)90138-l

[21]Samuel Rathmanner and Marcus Hutter. 2011. A Philosophical Treatise of Universal Induction. Entropy 13, 6 (June 2011), 1076–1136. DOI:https://doi.org/10.3390/e13061076

[22] Carlton M. Caves, Christopher A. Fuchs, and Rüdiger Schack. 2002. Quantum probabilities as Bayesian probabilities. Physical Review A (July 2002). DOI:https://doi.org/10.1103/physreva.65.022305

[23] Scott Aaronson. 2004. Is Quantum Mechanics An Island in Theoryspace. arXiv: Quantum Physics,arXiv: Quantum Physics (January 2004).

[24] Daniel S. Abrams and Seth Lloyd. 2002. Nonlinear Quantum Mechanics Implies Polynomial-Time Solution for NP-complete and #P Problems. Physical Review Letters (July 2002), 3992–3995. DOI:https://doi.org/10.1103/physrevlett.81.3992

[25] W. K. Wootters. 2002. Statistical distance and Hilbert space. Physical Review D (July 2002), 357–362. DOI:https://doi.org/10.1103/physrevd.23.357

[26] John Archibald Wheeler. 2018. Information, Physics, Quantum: The Search for Links. In Feynman and Computation. 309–336. DOI:https://doi.org/10.1201/9780429500459-19

[27] MichaelA. Nielsen and IsaacL. Chuang. 2011. Quantum Computation and Quantum Information: 10th Anniversary Edition. (January 2011).